\documentclass[runningheads]{llncs}
\usepackage{graphicx}
\usepackage{cite}
\usepackage{listings}
\usepackage{hyperref}
\usepackage{xcolor}

\usepackage{siunitx}

\DeclareSIUnit{\million}{\text{m}}

\usepackage{makeidx}  
\usepackage{graphicx}
\usepackage[footnote,draft,silent,nomargin]{fixme} 
\newcommand{\name}{TIB-arXiv}
\newcommand{\arxiv}{\emph{arXiv}}
\newcommand{\pdf}{PDF viewer}

\begin{document}
\title{\name{}: An Alternative Search Portal for the arXiv Pre-print Server}
\titlerunning{\name{}}
%
\author{Matthias Springstein\inst{1}\orcidID{0000-0002-6509-8534} \and
Huu Hung Nguyen\inst{1}\orcidID{0000-0001-7597-4908} \and
Anett Hoppe\inst{1}\orcidID{0000-0002-1452-9509} \and
Ralph Ewerth\inst{1,2}\orcidID{0000-0003-0918-6297}}
\authorrunning{M. Springstein et al.}
%
\institute{Leibniz Information Centre for Science and Technology (TIB), Hannover, Germany \and L3S Research Center, Leibniz Universit\"at Hannover, Germany\\
\email{\{matthias.springstein, hung.nguyen, anett.hoppe, ralph.ewerth\}@tib.eu}}
\maketitle              
\begin{abstract}
\arxiv{} is a popular pre-print server focusing on natural science disciplines (e.g. physics, computer science, quantitative biology). As a platform with focus on easy publishing services it does not provide enhanced search functionality -- but offers programming interfaces which allow external parties to add these services. This paper presents extensions of the open source framework \arxiv{} Sanity Preserver (SP). With respect to the original framework, it derestricts the topical focus and allows for text-based search and visualisation of all papers in \arxiv{}. To this end, all papers are stored in a unified back-end; the extension provides enhanced search and ranking facilities and allows the exploration of \arxiv{} papers by a novel user interface.
\end{abstract}

\section{Introduction}
\label{sec:intro}

For a scientist it is crucial to keep up to date in his field of research -- but this task is becoming increasingly difficult.  One reason for this problem is that the number of new publications per day increases dramatically. Another reason is that today's scientists have several ways to publish their article and the publication system is becoming more heterogeneous. As a result, scientists have to devote more and more time to find articles that are relevant for their research.

A good indicator of this trend is the \arxiv{} pre-print server. The number of articles in the repository has increased linearly over the last 25 years, with more than 10,000 articles per month in 2017~\cite{arxiv_stats_2017}. The platform's objective is to provide a good and easy-to-use publishing service, whereas it does particularly focus on user interface and search. Thus, functionalities for enhanced search and sorting are missing. Anyhow, its maintainers are open to external partners developing novel services on top of \arxiv{}'s existing infrastructure~\cite{rieger2016arxiv}.

This paper presents \name{}\footnote{\url{https://labs.tib.eu/arxiv}}, a web-based tool for enhanced search and exploration of publications in \arxiv{}. A comfortable interface and specifically developed retrieval and ranking methods enable scientists to easily keep track of the current development in their research field, while social and collaborative functionalities facilitate interactive research processes.

\section{Related Work}
\label{sec:relat}
\arxiv{} is a very popular source for academic literature search. Thus, web-based tools for its (more) efficient use already exist (see \cite{marra2017astrophysicists} for a survey). Most of the applications create a topic or user-specific news feed, but they differ in their presentation of the found articles: \emph{Arxiv Sanity Preserver}~\cite{arxiv_sanity} generates thumbnails, \emph{Arxivsorter}~\cite{arxiver} shows figures extracted from the papers. Some of them explore re-ranking techniques, considering for instance the number of tweets referring to the article~\cite{arxiv_sanity} or the authors' names~\cite{arxivsorter}.
A more global approach is presented by \emph{PaperScape}~\cite{paperscape}: The tool visualises the \arxiv{} dataset in form of a map. In concrete terms, each paper in \arxiv{} is visualized by a circle whose size means the number of citations and the position depends on the cited paper. 

Overall, all tools have one big disadvantage in common: They usually limit their topical scope to a single domain of interest (e.g., \emph{Arxiv Sanity Preserver} focuses on computer vision research). There is thus no real support for the cross-sectional search necessary for interdisciplinary search. Instead, the user would have to refer to different, domain-specific toolsets. Furthermore, the applications limit their scope to the re-ranking and representation of result lists. In contrast to that, we plan to provide an integrated platform for individualised re-ranking and search, and exploration of the document collection.


\section{\name{}: Focusing on search and user interface }
\label{sec:tool}

\begin{figure}
\centering
\includegraphics[width=\linewidth]{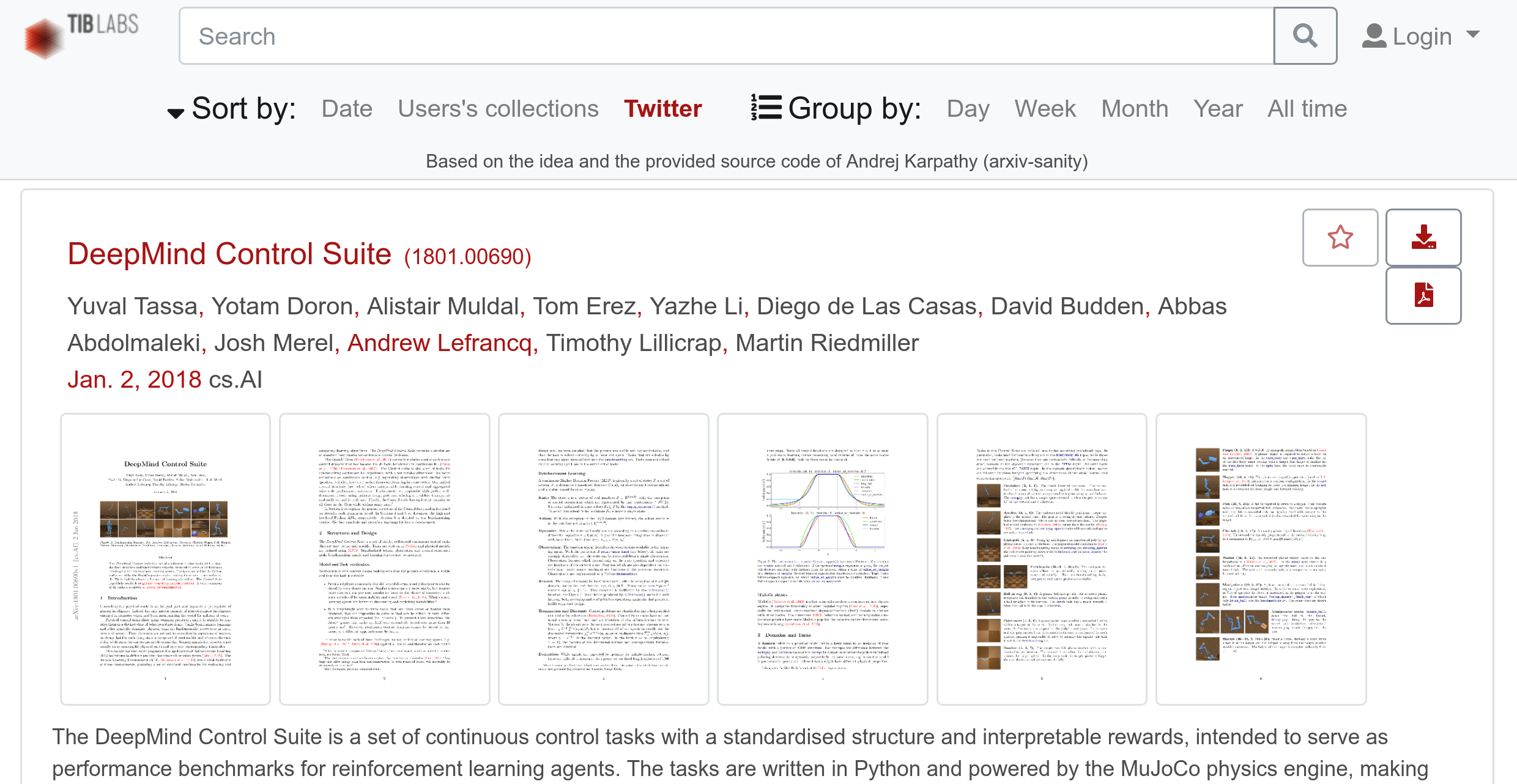}

\caption{Desktop version of \name{}}
\label{fig:overview}
\end{figure}


The web-based tool \name{} bundles the existing features of \arxiv{} and \emph{Arxiv Sanity Preserver}, and extends and improves them. It enables access to all \arxiv{} papers, offers efficient search and individualised ranking functionalities, provides an easy-to-use user interface which features an integrated PDF reader and additional visualisations.


\textbf{Data sources}: The \name{} project is based on the entire \arxiv{} data set and is synchronized daily. The website currently manages \SI{1.3}{\million} articles and preview images have been generated for more than half of them. The collected meta information contains the versions, title, authors and abstract of an article and is further extended by version metadata and a categorisation chosen by the author, based on the \arxiv{}-supported category set (e.g., cs.AI for computer science and artificial intelligence).

\textbf{Ranking}: \name{} offers several ways to rank the result list based on different criteria: 

\begin{itemize}
    \item \textbf{Date}: release date of the latest version
    \item \textbf{Twitter}: number of tweets that mention a certain paper
    \item \textbf{Collection}: number of copies in the individual users' collections
    \item \textbf{Relevance}: ranking based on the full-text search engine
\end{itemize}

Additionally, \name{} allows to restrict the list to results from a certain time span -- thus, the user can for instance explore papers which have been most popular on Twitter on a specific day or during the last month.

\textbf{Searching}:  The search is realised using Elasticsearch\footnote{\url{https://www.elastic.co/products/elasticsearch}} -- a full-text search engine based on the popular Lucene indexing and searching framework\footnote{\url{https://lucene.apache.org/}}. The index of \name{}  relies on the metadata provided by the \arxiv{} data set. Search can be limited to data fields selected by the user, for example, displaying only papers written by a certain author.
% ...

\begin{figure}
\centering
\includegraphics[width=\linewidth]{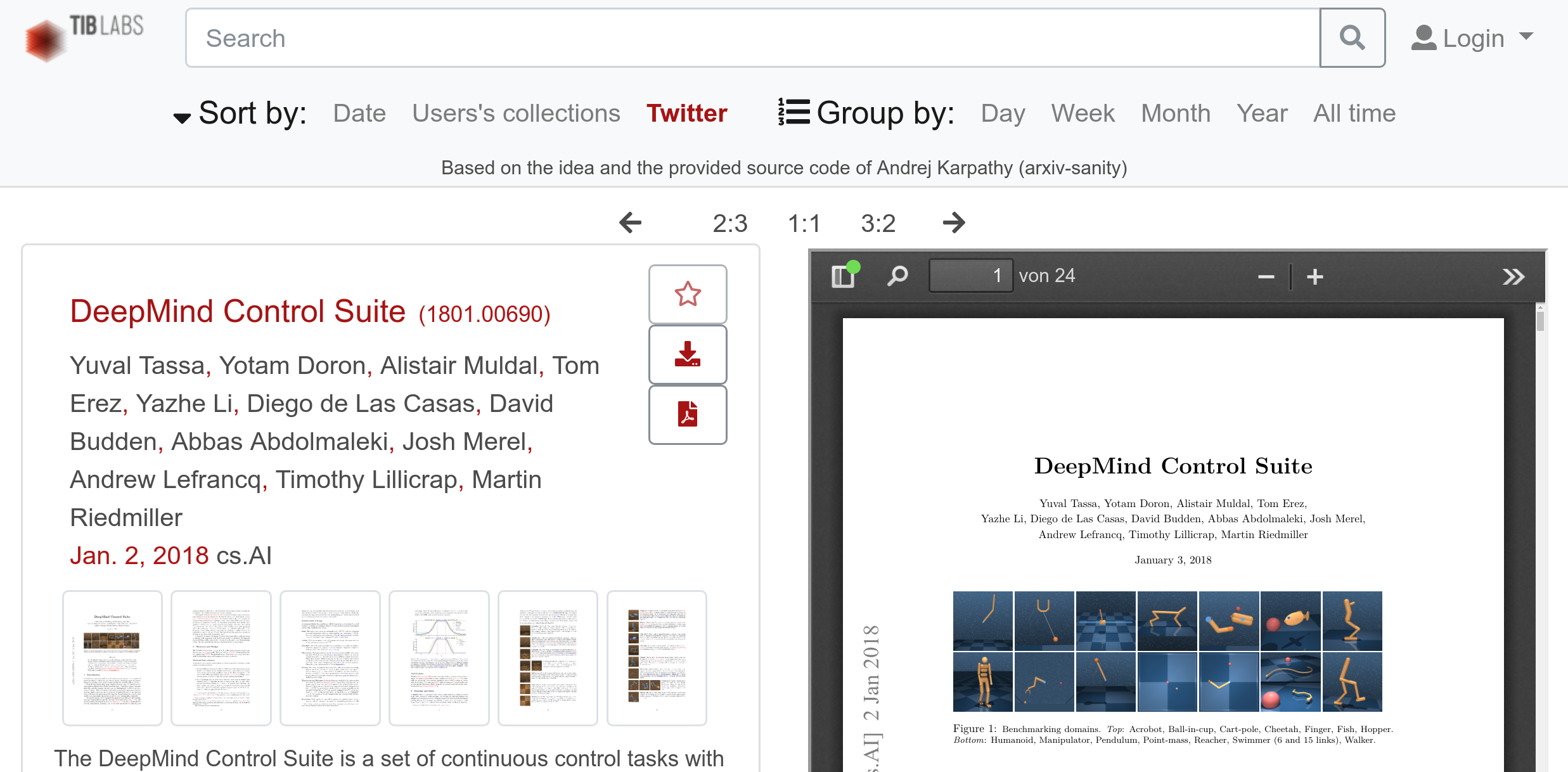}
\caption{Search results side by side with the embedded \pdf{}}
\label{fig:ui}

\end{figure}

\textbf{User Interface}:  \name{}'s user interface aims for clarity and responsiveness, and adapts to different display sizes, see Figure 1. Research articles are presented by their title, author list, research domains, thumbnails of their page content and their abstract.
Below the summary of the article, the user can find more detailed information about the selected document, including links to tweets that have mentioned this article.
Furthermore, the tool allows for direct interaction with relevant research articles -- they can be marked and stored for later usage, downloaded or read directly on the platform in an inline \pdf{}.
To handle different user preferences, \name{} allows to change the arrangement between the list view and the embedded \pdf{}. Finally, the interface can be displayed on mobile devices as well. The entire user interface is shown in Figure \ref{fig:ui}.




\section{Conclusions}
\label{sec:sf}


In this paper, we have described the current state of development of \name{}. It is aimed at overcoming some disadvantages of related tools and thus 1) enables access to all \arxiv{} papers, 2) offers efficient text-based search, 3) individualised ranking functionalities, and 4) provides a user interface with an integrated PDF reader, mobile access, and additional visualisations.

Anyhow, it is work in progress, and several further ideas are to be developed in the near future: The \pdf{} is to be more interactive -- interfacing with reference managers such as Mendeley could allow direct transfer and storage of citations and comments. Mass download services for individualised article packages (e.g. the works of a certain group, appeared in a certain time span, on a specific topic) could streamline scientific inquiry further. However, the focus on developing enhanced retrieval methods and strategies to alleviate the problem of scientific information overload.



\bibliographystyle{splncs04}
\bibliography{main}

\begin{thebibliography}{1}
\providecommand{\url}[1]{\texttt{#1}}
\providecommand{\urlprefix}{URL }
\providecommand{\doi}[1]{https://doi.org/#1}

\bibitem{arxiv_stats_2017}
arxiv.org::stats, \url{https://arxiv.org/help/stats/2017_by_area/index}

\bibitem{paperscape}
George, D.P., Knegjens, R.J.: Paperscape. \url{http://paperscape.org},
  accessed: 2018-04-11

\bibitem{arxiv_sanity}
Karpathy, A.: Arxiv sanity preserver. \url{http://www.arxiv-sanity.com},
  accessed: 2018-04-11

\bibitem{marra2017astrophysicists}
Marra, M.: Astrophysicists and physicists as creators of arxiv-based commenting
  resources for their research communities. an initial survey. Information
  Services \& Use  \textbf{37}(4),  371--387 (2017)

\bibitem{arxivsorter}
Ménard, B., Magué, J.P.: Arxivsorter. \url{http://www.arxivsorter.org},
  accessed: 2018-04-12

\bibitem{rieger2016arxiv}
Rieger, O.Y., Steinhart, G., Cooper, D.: arxiv@ 25: Key findings of a user
  survey. arXiv preprint arXiv:1607.08212  (2016)

\bibitem{arxiver}
Robert~Simpson, V.M., Hotan, A.: arxiver. \url{http://arxiver.net}, accessed:
  2018-04-11

\end{thebibliography}

\end{document}